\documentclass[aps,floatfix,amsmath,showpacs]{revtex4}
\usepackage{graphicx}
\usepackage{array}
\newcommand{\be}{\begin{equation}}
\newcommand{\ee}{\end{equation}}
\newcommand{\abe}{\begin{eqnarray}}
\newcommand{\aee}{\end{eqnarray}}
\newcommand{\no}{\noindent}
\begin{document}
\title{Chiral Vacuum Alignment and Spontaneous CP Violation \\by \\Four-Fermi Operators}\author{Tongu\c{c} Rador}
\email[]{tonguc.rador@boun.edu.tr}
\affiliation{Bo\~{g}azi\c{c}i University Department of Physics \\ 34342 Bebek, \.{I}stanbul, Turkey}

\begin{abstract} 
In models where there is a global chiral symmetry which spontaneously breaks to its vectorial subgroup, the introduction of an explicit symmetry breaking perturbation will define the true vacuum of the theory. This true vacuum is found via the minimization of the expectation value of the perturbing hamiltonian between different vacua as prescribed by Dashen. The procedure of finding the correct vacuum of the theory may result in the spontaneous breaking of CP symmetry even if one initially demands CP invariance on the perturbation. In this work we study in detail models where the perturbation is provided by four-Fermi operators. We present exact treatment for models with two fermion flavors and study the three flavor case in depth numerically. We show that after Dashen procedure is applied the solutions for the true vacuum fall in three classes with different CP breaking patterns. Critical transitions is possible between these classes as one varies the parameters of the perturbation. We rigorously show that at these transitions a pseudo-Goldstone boson mass vanishes.
We also advocate, and substantiate with numerical statistical analysis for various types of models, that if one imposes CP invariance on the perturbation before solving the vacuum alignment, the resulting vacuum structure will have a sizable probability for a light pseudo-Goldstone boson mass. That is a statistical variant of Peccei-Quinn mechanism can be speculated to operate. 
\end{abstract}

\pacs{12.60.Nz}
\maketitle

\section{Introduction}

Spontaneous breaking of continuous symmetries is an important part of science, occurring in a wide range of natural systems. When it happens the system admits continuously degenerate vacua. It is then important to pick a vacuum around which to define excited states of the system. For example, a collection of spins located on the sites of a given lattice will have minimum energy when all are parallel. But if this is all there is to it one can not prefer one direction to the other. However in the presence of a magnetic field, no matter how small, this direction is mandatorily the direction of the field defining the true vacuum of the theory. In this case, one can not ignore the presence of the perturbation and work with an arbitrary vacuum since this will lead to unstable excitations. The determination of this true vacuum state compatible with the perturbation is called vacuum alignment problem. In certain models, vacuum alignment may also spoil discrete symmetries that are ordinarily defined with respect to a vacuum, chosen to be a standard in the absence of a perturbation. Such a counter-intuitive result is shown to happen by Dashen \cite{dashen1,dashen2} for the combined symmetry of charge conjugation and parity (CP), within the context of mass perturbations of quarks. He showed that there are classes of mass perturbations that are initially naively CP conserving with respect to a standard vacuum yet after vacuum alignment spontaneously break CP symmetry.

In this work we will be studying vacuum alignment and spontaneous CP violation in theories where a global flavor chiral symmetry is spontaneously broken. We confine the perturbation that will explicitly break this chiral symmetry to the particular form of four-Fermi operators. Section II will present the model and the details of the Dashen program applied to it as well as the classification of CP breaking patterns. In Section III we present a mapping that simplifies the problem and allows a coherent exposition. Section IV is reserved to the phase transitions between different class of solutions as one varies parameters of the perturbing hamiltonian. We explicitly show that transitions between different class of solutions are accompanied with a vanishing pseudo-Goldstone boson mass. In Section V an exact treatment of the problem for two fermion flavors is presented. We also present a Monte-Carlo study where we have examined the statistical behavior of the pseudo-Goldstone boson masses and the CP breaking patterns for various types of models. As a result of this statistical study we also argue that an almost vanishing pseudo-Goldtone boson mass, that is fairly small compared to the largest one, is respectably probable. We argue that this result is closely connected to the fact that one imposes ab initio CP invariance on the perturbation. This effect bears a resemblance to the Peccei-Quinn mechanism \cite{pecqui1,pecqui2} where the anomaly parameter $\theta$, a measure of CP violation in strong interactions, is promoted to be a field and  is given an extra global U(1) symmetry that spontaneously breaks for  $\theta=0$. This consequently yields a Goldstone boson: the axion \cite{axion1,axion2}. However,  the Peccei-Quinn mechanism is exact and our analogy is valid only in the mean within a large collection of models. In the last section we discuss future directions to enlarge the present work.

\section{The Model}

Consider a collection of massless Dirac fermions $\Psi^{i}$ with $i=1,\cdots,N$. We assume that these fermions interact via a flavor blind, vectorial strong interaction described by a hamiltonian $H_{0}$. As is well known $H_{0}$ admits a chiral flavor symmetry $\rm{G_{C}=SU(N)_{L}\times SU(N)_{R}}$. Let us assume that the interaction becomes strong and confining at some scale $\Lambda_{T}$ thereby triggering the spontaneous breaking of $\rm{G_{C}}$ down to its vectorial subgroup $\rm{G_{V}=SU(N)_V}$, similar to the mechanism in QCD. The ground state of the theory is degenerate under this vectorial subgroup. This continuous degeneracy manifests itself as the appearance of massless excitations in the spectrum of the theory, namely the Goldstone bosons. The condensate characterizing this spontaneous breaking can be described as

\be{\label{eq:cond}}
\langle 0 \vert \bar{\Psi}_{L}^{i}\Psi^{j}_{R}\vert 0 \rangle=-\Delta_{T}\delta^{ij}
\ee

\no Here $\Delta_{T}\approx2\pi F^{3}$ and $F$ is the decay constant of the Goldstone bosons of the theory assuming a similarity with QCD \footnote{As an explicit example one can take $H_{0}$ as describing a scaled-up QCD type model.}. We also have $\Lambda_{T}\approx 4\pi F$. The state $\vert 0\rangle$ is called the {\em standard vacuum} and is invariant under $\rm{G_{V}}$. Using this fact one can also show that the equation defining the condensate is invariant under $\rm{G_{V}}$.

However (\ref{eq:cond}) is not invariant under the action of the full group $\rm{G_{C}}$ nor under the action of the co-set space $\rm{G_{C}/G_{V}}$ characterizing the Goldstone bosons. In this case we have

\be
\langle 0 \vert \bar{\psi}^{i}_{L}\psi^{j}_{R}\vert 0 \rangle=-\Delta_{T}W^{ij}
\ee

\no with $\psi_{L,R}=W_{L,R}\Psi_{L,R}$ and $W=W_{L}^{\dagger}W_{R}$, where $W$ can also be thought of as parameterizing the Goldstone boson fields in a non-linear sigma model approach. This is the active point of view. One can also see this 
as written with respect to another vacuum $\vert\Omega\rangle$

\be
\langle\Omega\vert \bar{\Psi}^{i}_{L}\Psi^{j}_{R}\vert \Omega \rangle=-\Delta_{T}W^{ij}
\ee

\noindent where $\vert\Omega\rangle=U(W)\vert 0\rangle$ and $U(W)$ is a unitary  operator acting on the collection of ground states and carrying one to another. This is the passive point of view.

Now, what happens if one introduces a perturbing hamiltonian $H_{1}$ on top of $H_{0}$? The general answer to this is that one has to apply the Dashen's procedure \cite{dashen1} which in principle says that not all hamiltonians are compatible with the standard vacuum. That is, not all hamiltonians admit the standard vacuum as the state which minimizes the energy. We then have the Dashen's theorem which says that the compatible vacuum state is defined by the state which minimizes the following

\be
E(W)=\langle\Omega(W)\vert H_{1} \vert\Omega(W)\rangle=\langle0\vert U^{\dagger}(W)H_{1}U(W)\vert 0\rangle
\ee

\noindent Of course one can see this equation from both points of view. The active one which means that one rotates the hamiltonian so that it becomes compatible with the standard vacuum, or the passive one which means that one finds the vacuum  among degenerate vacua \footnote{Of course the set of compatible vacua may be described by a single state or a set of states degenerate continuously or discretely.} which is compatible with the hamiltonian. In this work we will use the active point of view. In this point of view Dahsen's second theorem states that for all the hamiltonians compatible with the standard vacuum the vacuum energy is minimized at $W=I$. That is, defining the rotated hamiltonian 

\be
H'_{1}=U^{\dagger}(W)H_{1}U(W)
\ee

\noindent to be a compatible hamiltonian, one arrives at that

\be
E(\tilde{W})=\langle 0\vert U^{\dagger}(\tilde{W})H'_{1}U(\tilde{W})\vert 0\rangle
\ee

\noindent is minimized at $\tilde{W}=I$. We have $U(I)={\mathcal I}$, with ${\mathcal{I}}$ representing the identity operator. 

Let us now assume that the perturbation has the following form

\be{\label{eq:hprime}}
H_{1}=\Lambda_{ijkl}\bar{\Psi}^{i}_{L}\gamma^{\mu}\Psi^{j}_{L}\bar{\Psi}^{k}_{R}\gamma_{\mu}\Psi^{l}_{R}
\ee

\noindent where the summation convention is understood on repeated indices. We use this convention throughout this work. 

The emergence of a perturbation of the type (\ref{eq:hprime}) can for instance be motivated in technicolor theories (describing in our language $H_{0}$) enriched with extended technicolor which is itself a gauge theory (see for instance \cite{kentech1} for an introduction to technicolor and extended technicolor). Here to keep the analysis short we do not go into the details of motivating this any further, a new interaction
that couples the left-handed currents to right-handed ones is sufficient. If this is the case the object $\Lambda_{ijkl}$ will be proportional to the inverse mass matrix of the gauge bosons of the new interaction which presumably breaks at a higher energy scale than the scale at which $H_{0}$ creates a condensate \footnote{This is necessary because the perturbations which will explicitly break the chiral symmetry must really be {\em perturbations} on top of the spontaneous breaking. If the explicit breaking terms are more important than the mechanism triggering the spontaneous breaking there is no sense in talking about a symmetry at all}. Of course in such a scenario one would just as well have perturbations that couple two left(right)-handed currents, but these will have vanishing expectation values in any of the degenerate chiral vacua \footnote{The Dashen procedure, however, will change all these interactions accordingly when one rotates the full perturbing hamiltonian to become compatible with the standard vacuum}.

Without going into the details of the calculation the Dashen procedure applied to this problem means that one has to minimize the following quantity

\be
E(W)=-\Delta_{TT}\Lambda_{ijkl}W^{\dagger li}W^{jk}
\ee

\noindent subject to the fact that $W=W_{L}^{\dagger}W_{R}$ is a SU(N) matrix. In a QCD like theory $\Delta_{TT}\approx\Delta_{T}^{2}$, but from now on we will discard the dimensionful constants and
study with the appropriately normalized dimensionless objects. In this case the energy function simply becomes

\be{\label{eq:en1}}
E(W)=-\Lambda_{ijkl}W^{\dagger li}W^{jk}
\ee

\noindent where now $\Lambda$'s are dimensionless. Furthermore, since the perturbation explicitly breaks the chiral symmetry the Goldstone bosons will acquire masses. The mass matrix of these pseudo-Goldstone bosons (PGB), which is essentially the second derivative of the energy at the minimum, is given by the following

\be
M^{2}_{ab}=2\Lambda_{ijkl}\left[\left(\{\lambda_{a},\lambda_{b}\}W^{\dagger}\right)^{li}W^{jk}+W^{\dagger li}\left(W\{\lambda_{a},\lambda_{b}\}\right)^{jk}-2\left(\lambda_{a}W^{\dagger}\right)^{li}\left(W\lambda_{b}\right)^{jk}-2\left(\lambda_{b}W^{\dagger}\right)^{li}\left(W\lambda_{a}\right)^{jk}\right]
\ee

\noindent Here $\lambda_{a}$'s represent the generators of SU(N) in the fundamental representation, and $\{a,b\}\equiv ab+ba$ is the anti-commutator.

The minimization procedure applied to $E(W)$ does not determine $W_{L}$ and $W_{R}$ separately and hence the rotated hamiltonian is not uniquely defined. However this degree is due to the fact that the standard vacuum in invariant under the vectorial subgroup. One can, as a convention, take $W=W_{L}$ without loss of generality. In this case the rotated hamiltonian becomes

\be{\label{eq:roth1}}
{\Lambda_{W}}^{ijkl}=\Lambda^{i'j'kl}W^{\dagger ii'}W^{j'j}
\ee

\noindent In terms of this rotated hamiltonian the PGB mass matrix becomes

\be
M^{2\;ab}=2\Lambda_{W}^{ijkl}\left[\{\lambda^{a},\lambda^{b}\}_{li}\delta_{jk}+\delta_{ li}\{\lambda^{a},\lambda^{b}\}_{jk}-2\lambda^{a}_{li}\lambda^{b}_{jk}-2\lambda^{b}_{li}\lambda^{a}_{jk}\right]
\ee

\noindent{\bf Conditions on $\Lambda_{ijkl}$:} The hermiticity of the hamiltonian requires we impose the following

\be
\left(\Lambda^{ijkl}\right)^{*}=\Lambda^{jilk}
\ee

\noindent ensuring that $E(W)=E^{*}(W)$. One can also impose time-reversal invariance on the hamiltonian, which is equivalent to imposing CP invariance in view of the CPT theorem. This means that we must have

\be
\Lambda^{ijkl}\in\mathcal{R}
\ee

\noindent that is they are real. This invariance mandates

\be
E(W)=E(W^{*})
\ee

\noindent In terms of the rotated hamiltonian this will read

\be
\Lambda_{W^{*}}^{ijkl}=\left(\Lambda_{W}^{ijkl}\right)^{*}=\Lambda_{W}^{jilk}
\ee

Now, if $W\neq W^{*}$ CP will be spontaneously broken. The philosophy behind this idea goes back to Dashen's work, where he showed that one can have complex numbers in a theory even if initially one imposes reality via T invariance. The complex numbers arise as a result of the fact that one has to rotate the hamiltonian to become compatible with the standard vacuum. One can also see this from the passive point of view. One defines CP invariance in the standard vacuum, yet the hamiltonian is compatible with another vacuum for which the original CP transformation may not necessarily ensure invariance. Dashen showed this effect to occur in a very simple model of quark mass perturbation. Even though the mass matrix he proposed was real it nevertheless had a non-vanishing phase for the determinant. That is, albeit there ${\rm argdet(M_{q})=\pi}$,  the contribution to the chiral anomaly term was non-zero and consequently one has relevant complex phases (that is phases one can not rotate away via simple redefinitions of the fields) in the theory. This idea was promoted \cite{kentech2} to technicolor and extended technicolor to study CP violation via chiral vacuum alignment. The model was later studied in \cite{kenestiame} and to somewhat larger extent in \cite{kenadam1,kenadam2,kenadam3}. In this work we aim at enlarging the understanding of this model.

\subsection{Rational Phase Solutions}

In \cite{kenestiame} it was discovered that the minimization of (\ref{eq:en1}) gives three classes of solutions

\begin{itemize}
\item $W=W^{*}$, here $W$ is simply an orthogonal matrix, CP is not spontaneously broken.
\item $W\neq W^{*}$ with arbitrary complexity, CP is spontaneously broken arbitrarily.
\item $W\neq W^{*}$ where all the phases of $W$ are rational multiples of $\pi$, CP is spontaneously broken with a pattern.
\end{itemize}

\noindent In \cite{kenadam2} these solutions are called CP-conserving (CPC), CP-violating (CPV) and pseudo CP-conserving (PCP) respectively. We will follow this naming scheme. This classification is also valid for the rotated hamiltonian (\ref{eq:roth1}) as was shown in \cite{kenestiame}. That is, in a CPC phase the rotated hamiltonian is real, in a PCP phase it has phases that are rational powers of $\pi$ and in a CPV phase its phases are irrational multiples of $\pi$. There is an exception to this, if all of the complexity of $W$ comes as an overall phase the rotated hamiltonian will always be a CPC one. An overall phase containing all of the complexity means that one has $W=e^{i\chi}O$ with $O$ real and orthogonal and this will fix $\chi$ unambiguously depending on whether ${\rm det}O$ is $1$ or $-1$.

At this point we would like to mention that since the energy function is quadratic \footnote{In this work we use the word {\em quadratic} to emphasize that the energy function is proportional to $W_{il}W^{\dagger}_{jk}$. A general quadratic form will depend on ${\rm det}(W)$.} in $W$ every solution can be multiplied with a matrix $\exp(2\pi i k/N)$ for $k=0,1,\cdots,N-1$, to generate a new solution. In short every solution is  degenerate under $Z_{N}$, the center of $SU(N)$. With this remark we understand that if the complexity of $W$ is carried by an overall phase this phase must be and element of $Z_{N}$ or $\sqrt{Z_{N}}$.

The possibility for CPC and PCP phases is understood via the phase locking mechanism as explained in detail in \cite{kenestiame,kenadam2}. This mechanism is present because the energy function is quadratic in $W$. To review the idea let us rewrite the energy as follows

\be
E=-\Lambda_{ijkl}|W^{li}||W^{jk}|e^{i(\phi_{jk}-\phi_{li})}
\ee

\noindent If for a particular $ijkl$, $\Lambda_{ijkl}>0$ the contribution of this term to the total energy will be minimized if the phases of $W^{li}$ and $W^{jk}$ obey $\phi_{jk}=\phi_{li}$ or if $\Lambda_{ijkl}<0$ the contribution of this term to the total energy will be minimized if $\phi_{jk}=\phi_{li}+\pi$. This is the phase locking mechanism. However since $W$ is constrained to be an SU(N) matrix this cannot happen for all the elements, that is the phase locking may or may not be frustrated by the unitarity and unimodularity of $W$.
But when it is allowed this is the mechanism that triggers the solution to be in the  CPC or PCP phases. If the mechanism is not operational at all for a given model the solution will be a CPV one.

Let us consider the following parameterization 

\be
W=D_{1}W_{CKM}D_{2}
\ee 

\noindent where $D_{1}$ and $D_{2}$ are diagonal SU(N) matrices and $W_{CKM}$ is a Cabibbo-Kobayashi-Maskawa type matrix. This representation is called the Harari-Leurer representation \cite{harleur}. $W_{CKM}$ generally has phases which also infest the norms of its elements. That is not all the elements of $W_{CKM}$ have norms and phases that are independent of each other. This is an obstruction for phase locking. In fact in \cite{kenestiame} it was shown that for phase locking mechanism to occur one must have

\be{\label{eq:lock}}
W=D_{1} O D_{2}
\ee

\noindent where $O$ is an SO(N) matrix. In this case the phases of all the elements of $W$ are independent degrees than those constituting the norms of $W$, consequently phase locking can not be frustrated.

SU(N) representations are in general complex. However there are cases where the complex conjugate of an element may be obtained
by a conjugation with another element of the group. In this case one has

\be
W^{*}=g^{-1}Wg
\ee

\noindent and one says $W$ is a pseudo-real element. We would like to recall that since the energy function we are dealing with is quadratic in $W$ and thus that any solution to minimization is defined modulo $Z_{N}$, the above equation must actually read

\be{\label{eq:lockfin2}}
W^{*}=Z_{N}g^{-1}Wg
\ee

\noindent It is clear that for a generally complex unitary matrix this condition can not be fulfilled. However the special form of (\ref{eq:lock}) allows for such a relation between $W$ and $W^{*}$. The condition that has to be satisfied is 

\be
D_{1}^{*}OD_{2}^{*}=Z_{N}g^{-1}D_{1}OD_{2}g
\ee

\noindent meaning

\be
O=Z_{N}D_{1}g^{-1}D_{1}OD_{2}gD_{2}
\ee

\noindent If $O$ is a general SO(N) matrix, for the condition above to hold one must have

\begin{subequations}
\abe
g^{-1}&=&Z_{N}D_{1}^{*2}\\
g&=&D_{2}^{*2}
\aee
\end{subequations}

\noindent and one ends up with the following 

\be
\left(D_{1}D_{2}\right)^{2}=Z_{N}
\ee

\noindent which in turn yields

\be{\label{eq:lockfin}}
D_{1}D_{2}=K\exp(k\pi i/N)
\ee

\noindent Here $k=0,1,\cdots,N-1$ and $K$ is a diagonal real matrix with $K^{2}=1$, that is $K$ is diagonal with $\pm 1$ as entries. Of course we must remember that $D_{1}$ and $D_{2}$ are diagonal SU(N) matrices which imposes ${\rm det}(D_{1}D_{2})=1$.

This analysis shows the emergence of rational phase solutions as was shown in \cite{kenestiame,kenadam1,kenadam2}, albeit via means of numerical analysis. To get  insight into its meaning let us consider (\ref{eq:lockfin}) with $k=0$ for simplicity. If we take 

\be{\label{eq:lockfin3}}
\left(D_{1}D_{2}\right)^{2}=I\Longrightarrow D_{1}D_{2}=K
\ee

\noindent and consider the following

\be
D_{1}O_{a}D_{2}D_{1}O_{b}D_{2}\equiv D_{1}O_{c}D_{2}
\ee

\noindent one can show that defining a new product rule

\be
O_{a}\star O_{b}=O_{c}
\ee

\noindent with $O_{a}\star O_{b}=O_{a}D_{2}D_{1}O_{b}$, all the axioms of group theory are satisfied. That is the form $D_{1}O_{a}D_{2}$ with the product defined as mentioned constitutes a group. Therefor various forms of $D_{1}D_{2}$ satisfying the pseudo-reality condition (\ref{eq:lockfin2}) characterize different embeddings of O(N) elements  into SU(N) in the form of $D_{1}O_{a}D_{2}$.

However these embeddings should not in general be called subgroups since they have a group structure with respect to a different product rule than that of the group they are part of, namely  SU(N) with the product defined as the usual matrix multiplication. The only one among these embeddings that is a subgroup is the one for $K=I$, in which case one really has a O(N) subgroup. Furthermore embeddings with $K\neq I$ are not connected to the identity element of SU(N) since their identity with respect to the product rule $\star$ is the matrix $K$. Within the jargon of group theory these embeddings can be called {\em orbits}. In the present context we
call them O(N) type orbits.

Further insight and new conditions will arise if we study the complex conjugate of these orbits. That is if

\be
D_{1}O_{a}D_{2}
\ee 

\noindent with $D_{1}D_{2}=K$ characterizes an orbit so does

\be
D^{*}_{1}O_{a}D^{*}_{2}
\ee

In general one does not expect these two orbits to coincide. If they do however one must have

\be
D^{*}_{1}O_{a}D^{*}_{2}=D_{1}\tilde{O}_{a}D_{2}
\ee

\noindent meaning that 

\be
\tilde{O}_{a}=D_{1}^{* 2}O_{a}D_{2}^{* 2}
\ee

\noindent in view of the reality of orthogonal matrices this translates to

\begin{subequations}
\abe
D_{1}^{2}&=&K_{1}\\
D_{2}^{2}&=&K_{2}
\aee
\end{subequations}

\noindent Where $K_{1}$ and $K_{2}$ are $K$-type matrices mentioned before. Combining these with (\ref{eq:lockfin3}) we get the following

\begin{subequations}
\abe
D_{1}^{2}=D_{2}^{2}&=&K'\\
D_{1}D_{2}&=&K
\aee
\end{subequations}

\noindent where $K'$ is again a $K$-type matrix. From this it is clear that possible $\pi/2$ phases will enter the game regardless of what SU(N) we are dealing with.

So far we have dealt with the full O(N) matrices in $D_{1}OD_{2}$. But what happens if one actually has an $O(N-1)$ matrix as a solution? In this case the rational phase structure is further enriched. Let us go back to the pseudo-complexity condition

\be
O=Z_{N}D_{1}g^{-1}D_{1}OD_{2}gD_{2}
\ee

\noindent If O is an element of O(N-1) the diagonal matrices $D_{1}$ and $D_{2}$ are defined modulo $Z_{N-1}$ since this will commute with O and in reality we have the following

\be
O=Z_{N}Z_{N-1}D_{1}g^{-1}D_{1}OD_{2}gD_{2}
\ee

\noindent bringing more possibilities into the space of rational solutions. 

Our analysis so far means that  for SU(N), rational phase solutions yield a $W$ that has phases of the following form

\be
n_{0}\frac{\pi}{N}\pm n_{1}\frac{\pi}{N-1}\pm n_{2}\frac{\pi}{N-2}\pm \cdots\pm n_{N-1}\pi
\ee

\noindent where $n_{i}$ are natural numbers.

To summarize our results we restate that rational phase solutions fall into O(N) type orbits of SU(N) as described above and that the orbits have group structures. Only one of those orbits is an actual subgroup of SU(N) with respect to the product rule of SU(N), the others are just subsets with a group structure defined with respect to another product rule. The orbits are characterized with $K$ which in the most general case is a diagonal matrix and obeys $K^{2}=I$. Thus in general  one has a discrete set of orbits characterizing the rational phase solutions. In terms of our classification CPC solutions are the O(N) subgroup, PCP solutions are in the orbits with $K\neq I$ and CPV solutions are generic elements of SU(N). 

In short the main reason for the possibility of rational phase solutions is that the energy function we are minimizing does not care about the overall phase of $W$. This also gives rise to the phase locking mechanism as presented above.

As a final remark we would like to add that if a $\Lambda_{ijkl}$ gives a rational phase solution for a given $D_{1}$ and $D_{2}$ a small perturbation of the model is likely to remain in an embedding characterized with the same $D_{1}$ and $D_{2}$ (modulo the overall $Z_{N}$ ambiguity). That is under a small perturbation of the hamiltonian one is likely to remain in the same orbit. We will elaborate on this in section \ref{sec:trans}.

\subsubsection{ Physical implications of rational phase solutions:}

In the previous section we have exposed the mathematical existance of rational phase solutions. Their physical relevance will generally
depend on the particulars of the model. First of all in the interpretation of the spinors, what are those spinors? Second in the physical
nature of the strong interaction that, by growing strong at low energies, will spontaneously break the global chiral flavor symmetry. Finally on the meaning of the perturbation that will explicitly break the chiral symmetry.

One interpretation that is known to the author is that of works \cite{kenestiame},\cite{kenadam1}. In simplified terms, there the spinors were technifermions, the strong interaction is technicolor and the explicit symmetry breaking is achieved by extended technicolor. The technifermions were also coupling to the quarks via extended technicolor four-Fermi interactions and thus provide a mass term for quarks which will in essence contain all the information necessary for the low energy theory. The use of rational phases in that interpretation was to naturally explain the observed fact that in the quark sector strong CP violation is absent but weak CP violation is accomodated. Or to put it in honest terms, for the phenomenology of extended technicolor the CPV cases are bad since they are in conflict with no strong CP violation and the CPC phases are unable to introduce a weak CP violation (the phase in the CKM matrix of quarks). The rational phase solutions opens up an avenue to circumvent these phenomenological obstacles.

There could be various other physical cases where the rational phase solutions may prove to be interesting. For instance it is tempting to speculate that they may find applications in solid state phenomena such as fractional quantum hall effect. As the purpose of this
work is to lay out the general mathematical structure we will not pursue any of these motivations. 

\section{The Adjoint Map}

It is possible to decompose the object $\Lambda^{ijkl}$ into the following form

\be{\label{eq:ad1}}
\Lambda^{ijkl}=\Lambda^{ab}\lambda_{a}^{ij}\lambda_{b}^{kl}
\ee

\noindent Where $\lambda$'s represent the generators of SU(N) in the fundamental representation enriched by $\lambda_{0}=I/\sqrt{2N}$ with $I$ representing the identity matrix. So we have $N^{2}$ of them. The possibility for such a decomposition is clear since the number of independent elements of $\Lambda^{ijkl}$ if they are allowed to be complex is $2N^{4}$ which matches the number of independent entries in $\Lambda^{ab}$ if they are allowed to be complex. In fact one can easily show that the decomposition is invertible. That is, given a $\Lambda^{ijkl}$ one computes $\Lambda_{ab}$  as

\be
\Lambda_{ab}=4\Lambda^{i'j'k'l'}\lambda_{a}^{j'i'}\lambda_{b}^{l'k'}
\ee

\noindent which can be reinserted in (\ref{eq:ad1}) and using the well known identity 

\be{\label{eq:i1}}
\lambda_{a}^{ij}\lambda_{a}^{kl}=\frac{1}{2}\delta^{il}\delta^{jk}
\ee

\no one can show that the decomposition is one-to-one. We remind the reader that in (\ref{eq:i1}) $\lambda_{0}$ is included in the sum over the generators.

One can also understand the validity of  the mentioned decomposition by remembering that the hamiltonian consists of operators of the form

\be
\bar{\Psi}_{L}^{i}\gamma^{\mu}\Psi_{L}^{j}\bar{\Psi}_{R}^{k}\gamma^{\mu}\Psi_{R}^{l}
\ee

\no which transforms as the $\left({\rm N_{L}}^*\times {\rm N_{L}}\right)\times\left({\rm N_{R}}^{*}\times {\rm N_{R}}\right)$ representation of ${\rm SU(N)}_{{\rm L}}\times {\rm SU(N)}_{{\rm R}}$. This product decomposes as

\be{\label{rep}}
{\rm Ad}_{{\rm L}}\times {\rm Ad}_{{\rm R}} \oplus {\rm Ad}_{{\rm L}}\times {\rm 1}_{{\rm R}} \oplus {\rm 1}_{{\rm L}}\times {\rm Ad}_{{\rm R}} \oplus {\rm 1}_{{\rm L}}\times {\rm 1}_{{\rm R}}
\ee

\noindent where Ad denotes the adjoint representation of SU(N) which has dimension $N^{2}-1$. The first term in (\ref{rep}) represent the elements without $\lambda_{0}$ for the decomposition in (\ref{eq:ad1}). The second and third terms represent contributions from $\Lambda_{a0}$ and $\Lambda_{0a}$ and the last one from $\Lambda_{00}$ respectively.

In terms of this decomposition the vacuum energy becomes

\be
E=-\Lambda^{ab}\rm{Tr}\left[W^{\dagger}\lambda_{a}W\lambda_{b}\right]
\ee

\no from which we realize that the terms involving $\lambda_{0}$ are irrelevant. The terms in $\rm Ad_{L}\times 1_{R}$ and $\rm 1_{L}\times Ad_{R}$ do not contribute to energy at all. The term $\rm 1_{L}\times 1_{R}$ contributes only a constant which will be immaterial for the minimization procedure \footnote{Of course in the presence of gravity a constant addition to the vacuum energy becomes important and can not be neglected}. So we can from the outset ignore the effect of $\lambda_{0}$ and assume we only use the fundamental generators of SU(N) in (\ref{eq:ad1}) where the decomposition is defined \footnote{This seems to spoil the invertibility of the decomposition. But terms with $\lambda_{0}$ represents elements of $\Lambda^{ijkl}$ which are irrelevant from the outset.}. In view of the obvious connection we would like to call the decomposition mentioned above the adjoint map of the problem.

\subsection{The independent elements of $\Lambda$ which are relevant}

The hermiticity condition on $\Lambda^{ijkl}$ means

\be
(\Lambda^{jilk})^{*}=\Lambda^{ijkl}\Longrightarrow \Lambda_{ab}\in {\rm \bf{R}}
\ee

\noindent as can be easily checked from (\ref{eq:ad1}). So a generic model will have $(N^{2}-1)^{2}$ relevant parameters. 

When however time-reversal symmetry is assumed in the original hamiltonian, that is $\Lambda^{ijkl}\in {\rm \bf{R}}$, the parameters are further constrained. The algebra of SU(N) has $\rm{N}^{2}-1$ elements; the Cartan subalgebra consisting of real diagonal generators of cardinality $N-1$ and two sets consisting of real symmetric and pure imaginary anti-symmetric generators, both of which having $N(N-1)/2$ elements. In this language the combined effect of the hermiticity and reality conditions on  $\Lambda^{ijkl}$ simply means that $\Lambda^{ab}$ are real and do not have elements that mix real symmetric and pure imaginary anti-symmetric spaces of the generators. This means that in a suitable ordering of the generators $\lambda^{a}$, $\Lambda^{ab}$ has the 
following form

\[
\left(
{\begin{tabular}{c c c |c}
& & & \\
& $\Lambda_{RR}$ &  & 0\\
& & & \\
\hline 
& 0 &  & $\Lambda_{II}$

\end{tabular}}
\right)
\]

\noindent We will always assume such an ordering is made and use it in the remaining part of this work. The upper left corner represents the space of real symmetric generators and the lower right corner represents the pure imaginary anti-symmetric generators. We also assume that within the set of real symmetric generators, the generators of the Cartan subalgebra is placed first. 

Finally, from these considerations, it is clear that after the reality and hermiticity conditions are imposed on $\Lambda^{ijkl}$, 
the number of relevant degrees of freedom is
\[
\left[\frac{N(N-1)}{2}+N-1\right]^2+\left[\frac{N(N-1)}{2}\right]^2=\frac{N^{2}(N^{2}+1)}{2}-N(N+1)+1
\]

\noindent instead of the naive $N^{2}(N^{2}+1)/2$ disregarding the parts of $\Lambda^{ijkl}$ which does not contribute to the vacuum energy.

\subsection{The Formulation of the Problem in the Adjoint Map}

Let us remember that in the adjoint map the energy function reads

\be
E=-\Lambda^{ab}\rm{Tr}\left[W^{\dagger}\lambda_{a}W\lambda_{b}\right]
\ee

This form of the energy function has a special meaning. The operation $W^{\dagger}\lambda_{a}W$ induces an automorphism of the SU(N) algebra which does not change its structure. In fact one has

\begin{subequations}{\label{eq:trans}}
\abe
W^{\dagger}\lambda_{a}W &=& S_{ba}\lambda_{b}\\
S_{ba} &=& 2{\rm{Tr}}\left[W^{\dagger}\lambda_{a}W\lambda_{b}\right]
\aee
\end{subequations}

\noindent This means that  if $W=\exp(i\omega_{a}\lambda_{a})$ one has $S=\exp(i\Omega_{a}(\omega)T_{a})$ where $T_{a}$ is a generator in the adjoint representation of the algebra, that is $(T_{a})_{bc}=if_{abc}$. From these defining equations of $S$ it is easy to show the following

\begin{subequations}{\label{eq:propS}}
\abe
S_{aa'}S_{bb'}S_{cc'}f_{a'b'c'}&=&f_{abc} \\
S_{aa'}S_{bb'}S_{cc'}d_{a'b'c'}&=&d_{abc} \\
S_{aa'}S_{bb'}\delta_{a'b'}&=&\delta_{ab}
\aee
\end{subequations}

\noindent similar relations will hold for all the invariant tensors of the algebra, in particular it is clear from the above that $S$ must be an orthogonal matrix with unit determinant.

Written in this form, the energy function reads

\be
E=-\frac{1}{2}\Lambda_{ab}S_{ba}=-\frac{1}{2}{\rm{Tr}}\left[\Lambda S\right]
\ee

One can carry all the necessary formulas from the original setting to the adjoint map and in general they are much simplified. For example the rotated hamiltonian in the adjoint map can be computed from its definition in the original setting

\be
\Lambda_{W}^{ijkl}=\Lambda^{i'j'kl}W^{\dagger}_{ii'}W_{j'j}
\ee 

\noindent which results in

\be
\Lambda_{W}^{ijkl}=\Lambda^{ab}\left(W^{\dagger}\lambda_{a}W\right)^{ij}\lambda_{b}^{kl}=\Lambda^{ab}S_{bc}\lambda_{b}^{ij}\lambda_{c}^{kl}
\ee 

\noindent Thus in the adjoint map the rotated hamiltonian is represented as

\be
\Lambda_{S}\equiv\tilde{\Lambda}=\Lambda S
\ee

The equivalence of the two points of view is evident since $\Omega_{a}$ defining $S$ and $\omega_{a}$ defining $W$ are related. So one may look at $\Omega_{a}$ as defining $\omega_{a}(\Omega)$ and use the standard Dashen procedure in the adjoint map \footnote{Needless to say one can carry all the formulas in the original formulation to the adjoint map by brute force computation, using the decomposition in (\ref{eq:ad1}).}.

The extremization condition is simply,

\be
{\rm Tr}\left[\tilde{\Lambda} T_{a}\right]=0
\ee

\noindent We would like to emphasize that since $T_{a}$ are antisymmetric this is a condition on the antisymmetric part of $\tilde{\Lambda}$. One can show that the anti-symmetric part of the perturbation in the adjoint map represent the part  that breaks parity symmetry. {\bf So the extremization condition is a constraint on parity breaking} \footnote{Since we are using the active point of view, that is we rotate the perturbation not the vacuum, the definition of parity symmetry with respect to the standard vacuum is not altered. So the extremization condition is really trying to minimize parity breaking.} 

The pseudo-Goldstone boson mass matrix is given as

\be
M^{2}_{ab}=\frac{1}{2}{\rm Tr}\left[\tilde{\Lambda} (T_{a}T_{b}+T_{b}T_{a})\right]
\ee

\noindent The antisymmetric part of $\tilde{\Lambda}$ does not contribute to this (neither it does to the energy) so one can also write
is as

\be
M^{2}_{ab}={\rm Tr}\left[\tilde{\Lambda}T_{a}T_{b}\right]
\ee

from which we also get

\be
E=-\frac{1}{N}{\rm Tr}\left[M^{2}\right]
\ee

The usefulness of the adjoint map becomes clear if we perceive that a quadratic functional became a linear one and consequently the extremization condition as well as the PGB mass matrix take on a particularly simple form.

\subsection{$Z_{N}$ degeneracies and the adjoint map of $W^{*}$}

It is important to understand how the degeneracies of the original problem is changed after the adjoint map. If we remember the defining relation for $S$

\be
S_{ba} = 2{\rm{Tr}}\left[W^{\dagger}\lambda_{a}W\lambda_{b}\right]
\ee

\noindent we see that the $Z_{N}$ degeneracies disappear. That is if $W$ is mapped into $S$ then so is $Z_{N}W$.

As for the degeneracy $E(W)=E(W^{*})$, if $W$ is mapped to $S(W)$ then $W^{*}$ is mapped to

\be
S(W^{*})_{ba} = 2{\rm{Tr}}\left[W^{T}\lambda_{a}W^{*}\lambda_{b}\right]
\ee

By using the properties of the generators $\lambda_{a}$ one can show that if one has

\[
 S(W)=
\left(
{\begin{tabular}{c c c |c}
& & & \\
& $S_{RR}$ &  & $S_{RI}$\\
& & & \\
\hline 
& $S_{IR}$ &  & $S_{II}$
\end{tabular}}
\right)\;,
\]

then one has

\[ 
S(W^{*})=
\left(
{\begin{tabular}{c c c |c}
& & & \\
& $S_{RR}$ &  & $-S_{RI}$\\
& & & \\
\hline 
& $-S_{IR}$ &  & $S_{II}$
\end{tabular}}
\right)\;.
\]

The presence of a nonzero $S_{RI}$ and $S_{IR}$ means that the related $W$ is in general complex. If $S_{RI}=0$ and $S_{IR}=0$ the implication is that the rotated hamiltonian $\Lambda_{W}^{ijkl}$ is not a CPV one.

It is worth mentioning here that the effect of complex conjugation of $W$ can be represented in the adjoint map as a similarity transformation in $\rm{O(N^{2}-1)}$. That is, one has

\be
S(W^{*})=C^{-1}S(W)C
\ee

with 

\[ 
C=
\left(
{\begin{tabular}{c c c |c}
& & & \\
& I &  & 0\\
& & & \\
\hline 
& 0 &  & -I
\end{tabular}}
\right)
\]

An original hamiltonian yielding a $\Lambda$ such that $\Lambda_{RI}=0$ and $\Lambda_{IR}=0$ (meaning that $\Lambda^{ijkl}$ are real) will commute with this and so the fact that $S(W^{*})$ and $S(W)$ yielding the same energies will remain valid, as expected. This is again benefit of the use of the adjoint map, it transforms complex conjugation in the original framework to a simple form. But, of course, the matrix $C$ lives in $\rm{O(N^{2}-1)}$, outside the space where $S$ resides.

\subsubsection{Correspondance to the Non-Linear Sigma Model Language}

The adjoint map also makes a better connection to non-linear sigma model realization of the four-Fermi interaction we are studying. To make our formalism apparent in that language all we have to do is to remember that $W=W_{L}^{\dagger}W_{R}$ is really parameterizing the
Goldstone boson fields around the standard vacuum. Returning (for this discussion) the dimensionful parameters to their proper places we have $W=\exp\left[i\pi_{a}\lambda^{a}/F\right]\equiv\Sigma$ and the potential for the sigma field becomes

\be{\label{eq:nonlinpot}}
-\Delta_{T}^{2}\Lambda_{ab}\rm{Tr}\left[\Sigma^{\dagger}\lambda_{a}\Sigma\lambda_{b}\right]
\ee

\noindent The vacuum alignment procedure is equivalent to find the proper $\Sigma_{o}$ around which the pion tadpoles are vanishing  and pions have positive semi-definite masses. Such a term will only be relevant if it explicitly breaks the chiral flavor symmetry. Let us consider an explicit example of a LR symmetric model. In this case one has $\Lambda_{ijkl}=\mathcal{M}_{ij}\mathcal{M}_{kl}/M^{2}$ with $\mathcal{M}\in\mathcal{R}$ and $\mathcal{M}^{T}=\mathcal{M}$. The potential reads

\be
-\frac{\Delta_{T}^{2}}{M^{2}}\rm{Tr}\left[\Sigma^{\dagger}\mathcal{M}\Sigma\mathcal{M}\right]
\ee

\noindent and it is obvious that for a {\em generic} $\mathcal{M}$ the proper vacuum is given as $\Sigma_{o}=I$. Expanding around this we
see that the PGB mass matrix term is

\be
\frac{\Delta_{T}^{2}}{2M^{2}F^{2}}\pi^{a}\pi^{b}\rm{Tr}\left[\lambda_{a}\mathcal{M}\lambda_{b}\mathcal{M}\right]
\ee

\noindent We remind the reader that  $F$ is the decay constant of the goldstone bosons, $\Delta_{T}\approx 2\pi F^{3}$ is the value of the fermion condensate and $M^{2}$ is the scale of the four-Fermi interaction that explicitly breaks the chiral symmetry.

So in essence the adjoint map is close to the non-linear sigma model approach and is less cluttered than the $\Lambda_{ijkl}$ notation.
All one has to do is to study the behaviour of (\ref{eq:nonlinpot}) and this is what we are doing in essence. The transformation we have presented in (\ref{eq:trans}) (with the properties in (\ref{eq:propS}) understood as constraints) is simply rephrazing the non-linear sigma model in another language which provides technical ease. This is
valid since physical observables of non-linear sigma model is reparameterization invariant. 

\section{Digression on Transition Points}{\label{sec:trans}}

Consider the following functional

\be
E(x)=-\frac{1}{2}{\rm{Tr}}\left[\Lambda(x)S(x)\right]
\ee

\noindent where $S(x)$ is the matrix that minimizes the energy for a given $\Lambda(x)$. As we change $x$  we will vary $\Lambda(x)$, that is we follow a path in the space of hamiltonians. Thus the vacuum energy $E(x)$ and the corresponding $S(x)$ will change. 

Every now and then the excursion along $x$ traverses regions in the solution space where the corresponding rotated hamiltonian is real, arbitrarily complex or admits only rational phase complexity. That is, in the language presented before one may change between CPC, CPV or PCP phases. One may
traverse between any two regions. As we have discussed, the rational phase solutions CPC or PCP are related to {\em discrete} sets of orbits (different embeddings of O(N) into SU(N)), whereas a CPV phase lies in the bulk of the group space. Consequently, transitions between CPC and PCP (or between different PCP) phases are discrete changes in $W$ and hence in $S$. That is, $S$ changes discontinuously at the transition point. There is no such argument from the passages from CPC or PCP to the CPV phases, and one expects a continuous transition. That is, $S$ is changing continuously as one traverses the critical $x$. In this last case one expects a discontinuity in the derivative of $S$ because in CPV phases the solution is an arbitrary element of SU(N) but in CPC or PCP phases the solutions are in O(N)-type orbits so  one requires less number of generators to represent the group element and the number of generators representing a group element is related to its derivative. 

Another, albeit non-technical, way of seeing this structure is the following. CPV phases form the bulk of the solutions in the group space of SU(N) whereas CPC and PCP phases are different islands representing different embeddings of O(N) within the sea of CPV solutions. One may jump from island to island and this represents a discontinuous change in S. Whereas one may take a swim from an island or get ashore an island from the sea of CPV solutions. This second case represent a continuous change in $S$ since the sea and the shore are close in group space. But there nevertheless is a discrete change, one is either swimming or standing on an island. This
change represent itself in the first derivative of $S$ \footnote{Of course in this sea-island analogy one must remember that the dimensionalities of the islands and sea are different.}.

 The aim of this section is to show that at a continuous or discontinuous transition point at least one PGB mass vanishes. We will assume that $\Lambda(x)$ is a smooth function. That is, all of its derivatives exist. We follow a constructive scheme to prove the
summary proposition which is at the end of this section.

\vspace*{0.5cm}
{\bf Proposition 1}: The quantity ${\rm{Tr}}\left[\Lambda(x)\frac{dS(x)}{dx}\right]$ vanishes for all $x$. 
\vspace*{0.5cm}

This follows immediately from the fact that the extremization condition reads

\be
{\rm{Tr}}\left[\Lambda(x)S(x)T^{a}\right]=0
\ee

\noindent and the fact that the derivative of $S=\exp(i\Omega_{a}(x)T_{a})$ can always be represented in terms of the tangent space (the algebra),

\be{\label{eq:ds1}}
 \frac{dS(x)}{dx}=S(x)T_{a}G_{ab}(\Omega)\frac{d\Omega_{b}}{dx}
\ee

\noindent where $G$ is a general invertible matrix. As a corollary we have the following

\be
\frac{dE(x)}{dx}=-\frac{1}{2}{\rm{Tr}}\left[\frac{d\Lambda(x)}{dx}S(x)\right]
\ee

\vspace*{0.5cm}
{\bf Proposition 2}: $E(x)$ is a continuous function of $x$. 
\vspace*{0.5cm}

This follows from the smoothness of $\Lambda(x)$, for if $E(x)$ is assumed to have a discontinuity then $dE/dx$ must have a Dirac delta singularity. But if $\Lambda(x)$ is smooth, $dE/dx$ given by the corollary above is bounded from both above and below since $S(x)$ takes values from a compact space. This contradicts the assumption that $E(x)$ is discontinuous.

\vspace*{0.5cm}
{\bf Proposition 3}: The behavior of $S(x)$ near a point $x=0$ is determined by the PGB mass matrix at $x=0$, $S(0)$ and $d\Lambda/dx$ at $x=0$. .
\vspace*{0.5cm}

Let us consider a point near $x=0$ where $S(x)$ is given to be smooth. We can certainly expand $\Lambda(x)=\Lambda_{0}+x\Lambda_{1}$ since it is smooth. If $S(x)$ is smooth we can also expand it as $S(x)=S(0)(I+xs_{a}T_{a})$. The extremization condition is

\[
{\rm{Tr}}\left[(\Lambda_{0}+x\Lambda_{1})S(0)(I+xs_{a}T_{a})T_{b}\right]=0.
\]

\noindent and will give the following matching the like powers in an expansion up to first order in $x$, 

\begin{subequations}
\abe
{\rm{Tr}}\left[\Lambda_{0}S(0)T_{b}\right]&=&0\\
{\rm{Tr}}\left[\Lambda_{1}S(0)T_{a}\right]&=&{\rm{Tr}}\left[\Lambda_{0}S(0)T_{a}T_{b}\right]s_{b}
\aee
\end{subequations}

\noindent The first expression merely reemphasizes  the extremization condition for $x=0$. The second equation can be rewritten in terms
of the PGB matrix at $x=0$

\be
{\rm{Tr}}\left[\Lambda_{1}S(0)T_{a}\right]=M^{2}_{ab}(0)s_{b}
\ee

This is an equation for $s_{a}$ and thus determines how $S$ will change. It is clear that if the left hand side does not vanish (which is not expected for a generic $\Lambda_{1}$) there should be no zero eigenvalues in the PGB mass matrix, since one need to invert it to find $s_{a}$. If however the left hand side vanishes, we either have $s_{b}=0$, that is the solution for minimization does not change, or $s_{b}\neq0$. For the latter case there must be at least one zero eigenvalue and thus $s_{b}$ must be an element of the null space of the PGB mass matrix.
So the vanishing of ${\rm{Tr}}\left[\Lambda_{1}S(0)T_{b}\right]$ may trigger a degenerate minimum borrowing the jargon from Morse theory. It is worthwhile noting that this condition is the extremization equation for a rotated hamiltonian consisting of $\Lambda_{1}S(0)$ alone. Of course if $\Lambda_{1}$ is proportional to $\Lambda_{0}$ the solution must be $s_{a}=0$.  

Since in this corollary we have assumed that $S(x)$ is smooth we have to exercise a little more care in proving the existence of a vanishing PGB mass, the next two propositions deal with this.

\vspace*{0.5cm}
{\bf Proposition 4}: If $S(x)$ changes continuously at $x=0$ but its first derivative has a discontinuity, then at least one PGB mass must vanish at $x=0$.
\vspace*{0.5cm}

If we only have the continuity of $S(x)$ near $x=0$ we can expand it to the left and right as follows

\begin{subequations}
\abe
S(x>0)=S(0)(I+xs^{+}_{a}T_{a})\\
S(x<0)=S(0)(I+xs^{-}_{a}T_{a})
\aee
\end{subequations}

Then the extremization conditions to the left and right neighborhoods of $x=0$ will give the following

\begin{subequations}
\abe
{\rm{Tr}}\left[\Lambda_{0}S(0)T_{b}\right]&=&0\\
{\rm{Tr}}\left[\Lambda_{1}S(0)T_{b}\right]&=&M^{2}_{ab}(0)s^{+}_{b}\\
{\rm{Tr}}\left[\Lambda_{1}S(0)T_{b}\right]&=&M^{2}_{ab}(0)s^{-}_{b}
\aee
\end{subequations}

Again the first equation is just the fact that $x=0$ is an extremum, which is given from the outset. The last two equations can be used to give the following

\be
M^{2}_{ab}(0)(s^{+}_{b}-s^{-}_{b})=0
\ee

\noindent Since by assumption $s^{+}_{b}\neq s^{-}_{b}$ there must be at least one zero eigenvalue in the PGB mass matrix for 
a solution. It is clear from these considerations that both $s^{+}$ and $s^{-}$ must live in the null space for  if otherwise
they are to have any other part not in the null space one has to invert the PGB mass matrix to find it \footnote{If there is one vanishing PGB mass one does not necessarily have $s^{+}_{b}=s^{-}_{b}$, this would contradict the original assumption. In this case all is needed is that $s^{+}_{b}$ and $s^{-}_{b}$ must be proportional to each other.}.  This means that another condition that has to be satisfied is

\be
{\rm{Tr}}\left[\Lambda_{1}S(0)T_{b}\right]=0\;.
\ee

Finally we would like to remind that since the PGB mass matrix is given as ${\rm{Tr}}\left[\Lambda(x)S(x)T_{a}T_{b}\right]$ it changes continuously at a transition where $S$ is continuous. Thus the vanishing masses at the transition go to zero continuously.

If a discontinuity arises at the n'th derivative of $S(x)$, that is, if all the derivatives up to  and including the n-1'th derivative are continuous, the condition on the PGB masses does not differ from proposition 3. This type of transitions do not impose a vanishing PGB mass but puts restrictions on higher rank tensors since in this case one has to expand up to n'th order in $x$.

\vspace*{0.5cm}
{\bf Proposition 5}: If $S(x)$ changes discontinuously at $x=0$ then at least one PGB mass must vanish at $x=0$.
\vspace*{0.5cm}

If $S(x)$ changes discontinuously at $x=0$ we may pick 

\be{\label{eq:discc}}
S(x)=S^{-}(x)+\left[S^{+}(x)-S^{-}(x)\right]\Theta(x)
\ee

\noindent where without loss of generality we can assume that $S^{-}(x)$ and $S^{+}(x)$ are smooth at $x=0$, and thus can be expanded in $x$. As a result of our assumptions at $x=0$ there is a degeneracy in the minimization problem \footnote{This degeneracy is to be distinct from the $E(W)=E(W^{*})$ degeneracy. We also remind the reader that $Z_{N}$ degeneracies are absent in the adjoint map of the problem.}

\begin{subequations}{\label{eq:deg}}
\abe
{\rm{Tr}}\left[\Lambda(0)S^{+}(0)\right]&=&{\rm{Tr}}\left[\Lambda(0)S^{-}(0)\right]\\
{\rm{Tr}}\left[\Lambda(0)S^{+}(0)T_{a}\right]&=&0\\
{\rm{Tr}}\left[\Lambda(0)S^{-}(0)T_{a}\right]&=&0
\aee
\end{subequations}

Let us remember that the extremization condition holds for all $x$,

\be
{\rm{Tr}}\left[\Lambda(x)S(x)T_{a}\right]=0
\ee

\noindent thus we are allowed to take the derivative of this function and this gives

\be
{\rm{Tr}}\left[\frac{d\Lambda}{dx}S(x)T_{a}\right]+{\rm{Tr}}\left[\Lambda(x)\frac{dS(x)}{dx}T_{a}\right]=0
\ee

\noindent plugging in our ansatz in (\ref{eq:discc}) and using the conditions in (\ref{eq:deg}) one can show that

\begin{subequations}{\label{eq:deg2}}
\abe
{\rm{Tr}}\left[\frac{d\Lambda}{dx}(0)S^{+}(0)T_{a}\right]&=&0\\
{\rm{Tr}}\left[\frac{d\Lambda}{dx}(0)S^{-}(0)T_{a}\right]&=&0
\aee
\end{subequations}

\noindent which by virtue of proposition 3 means that we must have at least one vanishing PGB mass.

So a discontinuous transition in $S(x)$ is accompanied by a vanishing PGB mass. This makes a lot of sense if we remember that for a
small change in $x$ to induce a large change in $S(x)$ there has to be a direction in which the energy is not changing, which is another way of saying that a PGB mass vanishes. The situation is very similar to
the minimization of the function

\be
f(X,Y)=-\mu^{2}(X^{2}+Y^{2})+\lambda (X^{2}+Y^{2})^{2}+\epsilon X
\ee

\noindent with $\mu^{2}>0$ and $\lambda>0$. The true minimum lies along the $X$ axis for $\epsilon\neq 0$, let us call it $X_{o}$. Then, the sign of $X_{o}$ is correlated to the sign of $\epsilon$. When one traverses $\epsilon=0$ continuously $X_{o}$ changes discontinuously but at $\epsilon=0$ the minima is continuously degenerate along the angular direction, allowing the solution to move along it with zero energy cost.

\vspace*{0.2cm}
{\bf Corollary to Proposition 5}: Let us recall that the PGB mass-squared  matrix is given  as

\be
{\rm{Tr}}\left[\Lambda(x)S(x)T_{a}T_{b}\right]
\ee

\noindent At a discontinuous transition this matrix will presumably change discontinuously, thus the PGB masses will also change discontinuously. As we have shown at least one PGB mass must vanish, however we can not know if this zero happens as a discontinuous
drop in the mass or if it goes to zero continuously. All we know is that some masses may shift discontinuously. Nevertheless there is a condition on the shifts. Let us recall that the energy is alternatively given as 

\be
E(x)=-\frac{1}{N}{\rm{Tr}}\left[M^{2}(x)\right]
\ee

\noindent and that it is a continuous function regardless how $S$ changes. Thus at a discontinuous transition we have

\be
{\rm{Tr}}\left[M^{2}(0^{-})\right]={\rm{Tr}}\left[M^{2}(0^{+})\right]
\ee

\no So the relative shifts to the left and right balance. Not all masses can shift up or down, some must go up and some must go down
in such a way that the sum of the square of the masses remains the same.

Therefor we arrive at the corollary, if there is a discontinuous shift in the lowest PGB mass in one direction (that is either from $x$ negative to $x$ positive or vice-versa) it must go to zero continuously in the other direction. This follows from the fact that at least one PGB mass must vanish at a discontinuous transition.

\vspace*{0.2cm}
{\bf Summary Proposition:}
\vspace*{0.2cm}

At a transition point $x_{c}$ at least one PGB mass vanish. The necessary condition for $x_{c}$ to be a transition point is 

\be
{\rm{Tr}}\left[\frac{d\Lambda}{dx}(x_{c})S(x_{c})T_{a}\right]=0\;\;\;\;\;\;{\rm{and}}\;\;\;\;\;\frac{d\Lambda}{dx}(x_{c})\neq 0
\ee

This summarizes the previous results. From our line of reasoning it is not possible to prove that the above condition is also
a sufficient condition. Nor the above condition discriminates the transition as continuous or discontinuous. It should be mentioned that we assumed $\Lambda(x)$ to be smooth in this discussion and that it has a non-zero linear part in $x$ around $x_{c}$. If $\Lambda(x)$ is allowed to have discontinuities and/or singularities these will manifest themselves in $S(x)$ but these are to be considered as contrived transitions since they can be generated at ease.

At a continuous transition the PGB mass eigenstate will follow suite, that is, the eigenstate that hits zero has the same quantum numbers at both sides of the transition point. However at a discontinuous transition this is not expected since at different sides of the critical point the vacuum is discontinuously different. If on the other hand two eigenstates hits zero mass the situation will be more complicated since one will have to allow for an arbitrary superposition of the null space at the critical point regardless of wheter the transition is continuous or discontinuous.
\section{Exact Solution for SU(2)}

The case of SU(2) is much simplified since the adjoint formulation maps the problem in O(3)$\approx$SU(2), this is the reason an exact treatment is possible. The extremization condition reads

\be
\tilde{\Lambda}^{T}=\tilde{\Lambda}
\ee

\noindent That is the rotated hamiltonian $\tilde{\Lambda}=\Lambda S$ is symmetric. The reason for this simplification is that in SU(2) we have $(T_{a})_{bc}=i\epsilon_{abc}$, the fully anti-antisymmetric tensor. We have seen that the extremization conditions constraints parts of the perturbation that breaks parity symmetry. Here we realize its full effect: {\bf for the case of  SU(2) all rotated hamiltonians are parity conserving}. Furthermore the PGB mass-squared matrix is given as

\be
M^{2}={\rm Tr}(\tilde{\Lambda})I-\tilde{\Lambda}
\ee

\noindent where $I$ represents 3x3 identity matrix. The above means that $M^{2}$ and $\tilde{\Lambda}$ can be diagonalized simultaneously.

Now let us do a singular value decomposition of $\Lambda$ as follows

\be
\Lambda=L^{T}\Lambda_{d}R
\ee

\noindent where $L$ and $R$ are O(3) matrices and $\Lambda_{d}$ is a positive semi-definite diagonal matrix. With this the extremization condition reads

\be
\Lambda_{d}=K \Lambda_{d} K
\ee

\noindent with $K=RSL^{T}$. We also have $K^{T}K=I$ and ${\rm det}K={\rm sign(det}\Lambda)$. We assume without loss of generality that the determinant signature is carried by $L$, in such a way that finally one has ${\rm det}(S)=1$. 

After the singular value decomposition of $\Lambda$, the PGB mass-squared matrix becomes

\be
M^{2}={\rm Tr}(\Lambda_{d}K)I-L^{T}\Lambda_{d}K L
\ee 

\noindent We can define a new matrix $\tilde{M}^{2}=LM^{2}L^{T}$

\be
\tilde{M}^{2}={\rm Tr}(\Lambda_{d}K)I-\Lambda_{d}K
\ee

\noindent so that $\tilde{M}^{2}$ and $M^{2}$ are related via an orthogonal similarity transformation. Thus their eigenvalues are identical.

Now let us remember the extremization condition

\[
\Lambda_{d}=K\Lambda_{d}K
\]

\noindent this unambiguously means that

\be
\Lambda_{d}=K^{n}\Lambda_{d}K^{n}
\ee

\noindent for an arbitrary integer $n$, so at face value it seems there are countably infinite extrema. However let us recall that
$\Lambda^{d}$ is diagonal with positive entries. Without loss of generality we can assume $\Lambda_{d}={\rm diag}(a,b,c)$ with $a>b>c\geq0$ \footnote{This particular ordering of the singular values is not a must and our treatment of the problem can be carried out for any other ordering. We choose this form for simplicity in laying out the results}. If ${\rm det}(\Lambda)=0$ we can always choose
${\rm det}(L)={\rm det}(R)=1$ which would mean that ${\rm det}(K)=1$. We will deal with degeneracies in $\Lambda_{d}$ separately. The extremization condition along with this observation simply means that $K$ is diagonal and satisfies $K^{2}=I$. Therefor all the extrema fall into the following classes

\be{\label{eq:K1}}
{\rm det}(\Lambda)\geq 0\Longrightarrow K=
\left\{
{\begin{tabular}{c}
${\rm diag}(+1,+1,+1)$\\
${\rm diag}(+1,-1,-1)$\\
${\rm diag}(-1,+1,-1)$\\
${\rm diag}(-1,-1,+1)$\\
\end{tabular}}
\right.
\ee

\be{\label{eq:K2}}
{\rm det}(\Lambda)<0\Longrightarrow K=
\left\{
{\begin{tabular}{c}
${\rm diag}(+1,+1,-1)$\\
${\rm diag}(+1,-1,+1)$\\
${\rm diag}(-1,+1,+1)$\\
${\rm diag}(-1,-1,-1)$\\
\end{tabular}}
\right.
\ee

\begin{table}[t]

\begin{tabular}{|c|c|c|c|l|}
\hline
&&&&\\
det($\Lambda$)& K & E & PGB & {\rm Description} \\
\hline
\hline
$+$&$(+1,+1,+1)$&$-a-b-c$&$(+a+b,+a+c,+b+c)$& Minimum (Index 0 Extremum) \\ \hline
$+$&$(+1,-1,-1)$&$-a+b+c$&$(+a-c,+a-b,-b-c)$& Index 1 Extremum \\\hline
$+$&$(-1,+1,-1)$&$+a-b+c$&$(+b-c,-a+b,-a-c)$& Index 2 Extremum\\\hline
$+$&$(-1,-1,+1)$&$+a+b-c$&$(-a-b,-a+c,-b+c)$& Maximum (Index 3 Extremum)\\\hline\hline
$-$&$(+1,+1,-1)$&$-a-b+c$&$(+a+b,+a-c,+b-c)$& Minimum (Index 0 Extremum) \\\hline
$-$&$(+1,-1,+1)$&$-a+b-c$&$(+a+c,+a-b,-b+c)$& Index 1 Extremum \\\hline
$-$&$(-1,+1,+1)$&$+a-b-c$&$(+b+c,-a+c,-a+b)$& Index 2 Extremum\\\hline
$-$&$(-1,-1,-1)$&$+a+b+c$&$(-a-b,-a-c,-b-c)$& Maximum (Index 3 Extremum)\\
\hline
\end{tabular}
\caption{The extrema structure for $\Lambda_{d}={\rm diag}(a,b,c)$ with $a>b>c\geq0$.}
\label{table1}
\end{table}

\noindent The fact that we are using a different set of $K$'s for different signatures of $\det(\Lambda)$ is only to have a clear exposition. The careful reader might have already noticed that the set of $K$'s for $\det(\Lambda)<0$ are just the set of $K$'s for $\det(\Lambda)>0$ multiplied with $-I$. In fact, for SU(2) case, a negative determinant $\Lambda$ can be written as $-I\Lambda'$ where the singular value structure of $\Lambda'$ is the same as that of $\Lambda$. This allows one to use only one set of $K$'s, namely the one we use for $\det(\Lambda)>0$, but in this approach the maximum in one case is the minimum for the other. In view of this fact which may lead to unnecessary confusion we did not follow that line of exposition. 

Armed with (\ref{eq:K1}) and (\ref{eq:K2}) we now list the energies and associated PGB mass-squared values for {\em all} the extrema in TABLE \ref{table1}. In the table we have also indicated the description of the extrema in terms of the language of Morse theory, where the index of a non-degenerate (that is no vanishing PBG masses in our language) extremum is given by the number of negative eigenvalues of the Hessian (simply the PGB mass matrix in our terms). We would also like to remind the reader that, as a check, the number of extrema
satisfies the Morse inequalities. That is, the number of extrema of index $i$, is larger or equal to the $i$'th Betti number of the manifold. For SU(2)$\sim$S$^{3}$ the Betti numbers are $1,0,0,1$ and the number of extrema of our function are $1,1,1,1$ which satisfies the inequality as they should. How can we make use of Morse inequalities? For instance for  SU(3) the Betti numbers are $1,0,0,1,0,1,0,0,1$. These mean, apart from stating that a minimum and a maximum must occur (which is intuitive anyways since SU(3) is compact), that there must be at least one extrema with index 3 and 5. That is, extrema with 3 and 5 negative PGB masses respectively. Other than this, Morse inequalities, unfortunately are not too useful for an explicit solution. For completeness we remind the reader that the Betti numbers of SU(N) are given by the coefficients of the respective power in the Poincar\'e polynomial

\be
P_{SU(N)}(t)=(1+t^{3})(1+t^{5})\cdots(1+t^{2N-1})
\ee

\noindent{\bf Degeneracies in $\Lambda_{d}$}:
\vspace*{0.2cm}
Let us remember that the extremization condition is

\[
\Lambda_{d}=K\Lambda_{d}K
\]

\noindent If there are degeneracies in $\Lambda_{d}$ the behavior of the energy and the PGB masses can still be read from TABLE \ref{table1}. However the solution set for $K$ is larger. Let us start with an example. Assume that $\Lambda_{d}={\rm diag}(a,a,c)$, then  the possible solutions to the extremum condition are

\be
{\rm det}(\Lambda)\geq 0\Longrightarrow K=
\left\{
{\begin{tabular}{c}
${\rm diag}(+1,+1,+1)$\\
$\kappa_{+}(\theta)$\\
${\rm diag}(-1,-1,+1)$\\
\end{tabular}}
\right.
\ee

\be
{\rm det}(\Lambda)<0\Longrightarrow K=
\left\{
{\begin{tabular}{c}
${\rm diag}(+1,+1,-1)$\\
$\kappa_{-}(\theta)$\\
${\rm diag}(-1,-1,-1)$\\
\end{tabular}}
\right.
\ee

\noindent where we have 

\be
\kappa_{\pm}(\theta)=\left({\begin{tabular}{c c c}
$\cos\theta$& $\sin\theta$& $0$ \\
$\sin\theta$& $-\cos\theta$&  $0$\\
$0$& $0$& $\pm 1$ \\
\end{tabular}}\right)
\ee

\begin{table}
\begin{tabular}{|c||c|c|c|l|}
\hline
$det(\Lambda)$ & $K$ & $E$ & $PGB$ & {\rm Description} \\
\hline
\hline
$+$&$(+1;+1;+1)$    &$-2a-c$ & $(+2a,+a+c,+a+c)$ & Minimum (Index 0 Extremum) \\\hline
$+$&$\kappa_{+}(\theta)$ &$  +c $ & $(+a-c,\;0\;,-a-c)$ & Degenerate Critical Point \\\hline
$+$&$(-1;-1;+1)$    &$+2a-c$ & $(-a+c,-a+c,-2a)$ & Maximum (Index 3 Extremum)\\\hline\hline
$-$&$(+1;+1;-1)$    &$-2a+c$ & $(+2a,+a-c,+a-c)$ & Minimum (Index 0 Extremum) \\\hline
$-$&$\kappa_{-}(\theta)$ &$-c$    & $(+a+c,\;0\;,-a+c)$ & Degenerate Critical Point \\\hline
$-$&$(-1;-1;-1)$    &$+2a+c$ & $(-a-c,-a-c,-2a)$ & Maximum (Index 3 Extremum)\\\hline
\end{tabular}
\caption{The extrema structure for $\Lambda_{d}={\rm diag}(a,a,c)$ with $a>c\geq0$.}
\label{table2}
\end{table}

\begin{table}
\begin{tabular}{|c||c|c|c|l|}
\hline
$det(\Lambda)$ & $K$ & $E$ & $PGB$ & {\rm Description} \\
\hline
\hline
$+$&$(+1,+1,+1)$            &$-a-2b$ & $(+a+b,+a+b,+2b)$ & Minimum (Index 0 Extremum) \\\hline
$+$&$(+1,-1,-1)$            &$-a+2b$ & $(+a-b,+a-b,-2b)$ & Index 1 Extremum \\\hline
$+$&$\tilde{\kappa}_{+}(\theta)$ &$+a$    & $(\;0\;,-a+b,-a-b)$ & Degenerate Maxima\\\hline\hline
$-$&$\tilde{\kappa}_{-}(\theta)$ &$-a$    & $(+a+b,+a-b,\;0\;)$ & {\bf Degenerate Minima} \\\hline
$-$&$(-1,+1,+1)$            &$+a-2b$ & $(+2b,-a+b,-a+b)$ & Index 1 Extremum \\\hline
$-$&$(-1,-1,-1)$            &$+a+2b$ & $(-2b,-a-b,-a-b)$ & Maximum (Index 3 Extremum)\\\hline
\end{tabular}
\caption{The extrema structure for $\Lambda_{d}={\rm diag}(a,b,b)$ with $a>b\geq0$.}
\label{table3}
\end{table}

\begin{table}
\begin{tabular}{|c||c|c|c|l|}
\hline
$det(\Lambda)$ & $K$ & $E$ & $PGB$ & {\rm Description} \\
\hline
\hline
$+$&$(+1,+1,+1)$            &$-3a$   & $(+2a,+2a,+2a)$ & Minimum (Index 0 Extremum) \\\hline
$+$&$K_{+}(\theta,\phi)$    &$+a$    & $(\;0\;,\;0\;,-2a)$ & Degenerate Maxima\\\hline\hline
$-$&$K_{-}(\theta,\phi)$     &$-a$    & $(\;0\;,\;0\;,2a)$ & {\bf Degenerate Minima} \\\hline
$-$&$(-1,-1,-1)$            &$+3a$   & $(-2a,-2a,-2a)$ & Maximum (Index 3 Extremum)\\\hline
\end{tabular}
\caption{The extrema structure for $\Lambda_{d}={\rm diag}(a,a,a)$ with $a>0$.}
\label{table4}
\end{table}

\noindent These matrices obey $\kappa_{\pm}^{2}=I$ and are called reflection matrices. The results for the degenerate cases are summarized 
in TABLES \ref{table2}, \ref{table3} and \ref{table4}. The matrices $\tilde{\kappa}_{\pm}(\theta)$ and $K_{\pm}(\theta,\phi)$ referred in those tables are given
as

\be
\tilde{\kappa}_{\pm}(\theta)=\left({\begin{tabular}{c c c}
$\cos\theta$& $0$ &$\sin\theta$ \\
$0$ & $\pm 1$ &  $0$\\
$\sin\theta$& $0$& $-\cos\theta$ \\
\end{tabular}}\right)
\ee

\noindent and

\be
K_{\pm}(\theta)=\pm\left({\begin{tabular}{c c c}
$-1+2 \cos^{2}\theta\sin^{2}\phi$ & $\cos^{2}\theta\sin2\phi$& $\sin2\theta\sin\phi$\\
$\cos^{2}\theta\sin2\phi$ & $-1 +2\cos^{2}\theta\cos^{2}\phi$ & $ \sin2\theta\cos\phi$\\
$\sin2\theta\sin\phi$ & $\sin2\theta\cos\phi$ & $-1+2\sin^{2}\theta$\\
\end{tabular}}\right)
\ee

\subsection{Summary of Results for SU(2)}

It is clear from our treatment of the case of SU(2) that {\bf a degenerate minima occurs if the two lowest singular values of $\Lambda$ are degenerate and if determinant of $\Lambda$ is negative}. This condition is both necessary and sufficient. Now, what will happen, as we have discussed  previously, if one has a $\Lambda(x)$ and changes $x$ to have an excursion in the solution space? The condition for a degenerate minima may
be approached either by having the two lowest singular values to be the same and have $\det(\Lambda)$ crosses zero from positive to negative values, or having $\det(\Lambda)<0$ and force the two lowest singular values approach each other, or both of these mechanisms. In all of these ways if $\Lambda(x)$ has a non-vanishing linear part in $x$ around $x_{c}$ there will be a crossing from $K={\rm diag}(+1,+1,-1)$ to $K={\rm diag}(+1,-1,+1)$. Thus {\bf if $\Lambda(x)$ has a non-vanishing linear part in $x$ around $x_{c}$ the transition is a discontinuous transition in $S$}.

For SU(2) the rotated hamiltonian $\Lambda_{W}^{ijkl}$ is always real. The reason for this is that
after hermiticity and T-invariance is imposed, any hamiltonian in the adjoint map has the following form

\[
\Lambda=
\left(
{\begin{tabular}{c c c}
A & B & 0 \\
C & D & 0\\
0 & 0 & E \\
\end{tabular}}
\right)
\]

\noindent Where, in our ordering of the generators, the upper-left block represents the real-symmetric generators. This matrix can be diagonalized within the space of the real generators, that is $L$ and $R$ matrices diagonalizing $\Lambda$ do not mix real-symmetric and pure-imaginary anti-symmetric spaces of generators. Consequently the resulting $S$ will not mix these either. As a remark we also state that under these conditions, $L$ and $R$ will carry a single rotation angle each. The reason for this is that there is only one pure imaginary generator, a circumstance related to the fact that all representations of SU(2) are pseudo-real, that is a complex conjugation operator exists within the algebra \footnote{In the usual ordering of the generators this is given as $i\sigma_{2}$.}. Of course if one relaxes the T-invariance condition and consequently allow $\Lambda$ to be  a generic real 3x3 matrix one will have complex rotated hamiltonians $\Lambda_{W}^{ijkl}$.

In the previous paragraph we do not mean that the corresponding $W$ for a given $S$ will always be real. One can show using the $K$'s we have found and the structure of $L$ and $R$ following from the discussion above that $W$ is one of the following

\be{\label{eq:Ws}}
\left( {\begin{tabular}{cc} $\cos\theta$ & $\sin\theta$ \\ $-\sin\theta$ & $\cos\theta$ \end{tabular}}\right)\;\; , \;\;
\left( {\begin{tabular}{cc} $i\cos\phi$ & $i\sin\phi$ \\ $i\sin\phi$ & $-i\cos\phi$ \end{tabular}}\right)\;\; , \;\;
\left( {\begin{tabular}{cc} $-i\sin\phi$ & $i\cos\phi$ \\ $i\cos\phi$ & $i\sin\phi$ \end{tabular}}\right)\;\; , \;\;
\left( {\begin{tabular}{cc} $-\sin\theta$ & $\cos\theta$ \\ $-\cos\theta$ & $-\sin\theta$ \end{tabular}}\right)
\ee

\noindent here $\theta=(s+t)/2$ and $\phi=(s-t)/2$ where $s$ and $t$ represent the rotation angles in $L$ and $R$ respectively.

As can be seen from (\ref{eq:Ws}) the complexity of $W$ comes as an overall factor and consequently the rotated hamiltonian $\Lambda_{W}^{ijkl}$ will be real. So the solutions are always in the CPC phase. We would like to remind the reader that this happens because the energy
function we are minimizing does not care about the overall phase of $W$. Of course if there are other perturbations such as a mass term for the spinor these conclusions will change since the solution space will be altered.

\subsection{Frequency of Light PGB Masses}{\label{subsec1}}

Recently there has been considerable attention on models where the Higgs particle is a naturally light goldstone boson \cite{little1, little2, little3, little4}. This possibility is also advocated in within the context of the problem we are dealing with \cite{kenadam3}. So the question is: assuming we are sampling models from a random set what is the probability of having a light PGB mass? 

Our treatment of the problem for SU(2) is exact and in fact it is applicable not only for originally T-invariant hamiltonians but also for generic cases where $\Lambda$ is a 3x3 real matrix. As a consequence of this exact treatment the possibility for a vanishing PGB mass can be analyzed analytically. Even though this can be carried out  it would be very cumbersome, so we resort to  Monte-Carlo methods.

We present our findings for the following random samples. Models of type I are full hamiltonians where T-invariance is {\em \bf{not}} imposed, that is they are 3x3 matrices. Models of type II are full hamiltonians where T-invariance is imposed, that is they are block diagonal as discussed before. Models of type III are models where $\Lambda$ is diagonal and hence automatically T-invariant, thus one can see this type as constrained type II where off-diagonal entries are taken to be zero from the outset, thereby introducing the element of hierarchy in $\Lambda$. For each of these types we present three flavors. Flavor A where the elements are sampled from $\left[0,1\right]$, flavor B where elements are sampled from $\left[-1,0\right]$ and flavor C where elements are taken from $\left[-1,1\right]$.

\begin{figure}[t]
 \centering
 \includegraphics[scale=.75]{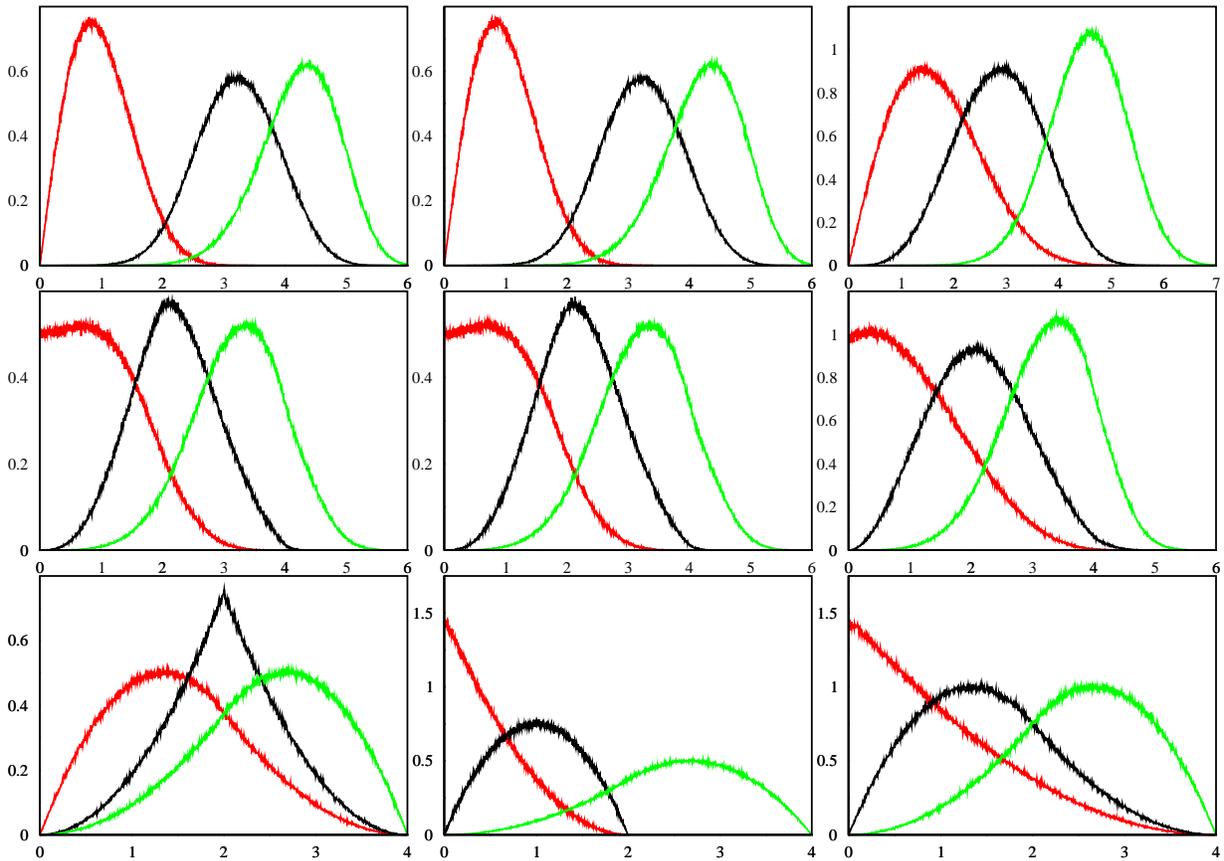}
 \caption{Normalized PGB mass-squared distributions for models discussed in the text. The columns and rows represent model type and flavor respectively. For example the rightmost graph in the middle row represent results model MODEL IIC. It is self evident which plot represents which PGB mass in a given graph but for clarity we picked red, black and green to represent the lightest, middle and heaviest PGB mass. The fluctuations in the graphs are representative of the finite number of samples we have taken for each model.}
 \label{fig:su2}
\end{figure}

For all these types and flavors we have sampled $3.2\;10^{6}$ models and solved the minimization problem numerically. In numerical minimization algorithms one can not be sure that the computed minimum is a global one. However, for SU(2), our exact treatment shows that modulo  $Z_{N=2}$ degeneracies the minimum is unique \footnote{Except of course when it is accompanied with a vanishing PGB mass} and thus the minimum found numerically is the global minimum. The normalized probability distributions of the PGB masses are in Fig.~\ref{fig:su2}.

Models IA and IB have the same PGB distributions, this is clear since sampling space of flavor B is just the negative of sampling space of flavor A. Since minimum and maximum structure of the solutions in SU(2) are symmetric this is expected. This behavior is also present in models IIA and IIB. However this symmetry does not exist between models IIIA and IIIB since in IIIA $\det(\Lambda)>0$ by construction and a PGB mass never vanishes. 

What is remarkable is that for models of type II and III there is considerable probability for the lightest PGB mass to be near zero.
This behavior is more emphasized for type III (except for IIIA where it is impossible to have a vanishing PGB mass from the outset).
This is to be contrasted to the behavior of models of type I  where T-invariance is not imposed. It is understandable that when one has less number of degrees of freedom it becomes more probable for those parameters to
contrive  so that the condition for a light PGB mass is satisfied.

We know from Morse lemma that functions with non-degenerate critical points are dense within the space of functions. That is, in our language, we should expect a generic model to have no zero PGB masses. In fact we observe this lemma; the probability for an
exactly vanishing PGB mass is zero. Nevertheless  the probability for a low PGB mass is considerable.
What is meant by a {\em truly generic} function in Morse theory translates to our study as a model where T-invariance is not a priori imposed. So the key element in the occurrence of light PGB masses is the assumption of T-invariance since this constrains the models. Constraining models further such as going from models of type II to type III makes the light PGB masses more likely. This constraint (vanishing of some elements) introduces a hierarchy in the explicit symmetry breaking hamiltonian.

The horizontal axis figures \ref{fig:su2} and \ref{fig:su3} represent the PGB mass squared value and the vertical axis is dimensionless since it represents the probability distribution. In terms of dimensionfull parameters the x axis is normalized in terms of $\Delta_{TT}/(M^{2}F^{2})$. The scale $M$ of the explicit perturbation must not interfere with the scale $4\pi F$ of the spontaneous symmetry breaking otherwise it can not be called a perturbation. Thus it would be desirable that the largest PGB mass is at most of order $F$, the decay constant of the Goldstone bosons. For instance for the case of SU(3) which we will present below we find that about \% 14 of the models have a PGB mass below $0.1 F$ for models in the set IIIB.

\section{Digression on SU(3)}{\label{sec:killtechni}}

Unfortunately an exact treatment of SU(3) case is not readily available. The adjoint map transforms the problem into O(8) matrices and a singular value decomposition of $\Lambda$ is not available as was in the case of SU(2). Even for an originally diagonal $\Lambda$ the analysis is difficult and we reserve it for future work. For certain simple cases an exact treatment could be possible but we do not pursue this here either. In SU(3) the interesting generalization is that the rotated hamiltonian can have rational phases and thus PCP solutions are possible. In this section we present our findings after the same Monte-Carlo analysis we did for SU(2). Since we do not
have an exact solution for SU(3) we can not say that the minimum is always unique and consequently the honest statement is that the numerical study is dealing with local minima.

\begin{figure}[t]
 \centering
 \includegraphics[scale=.75]{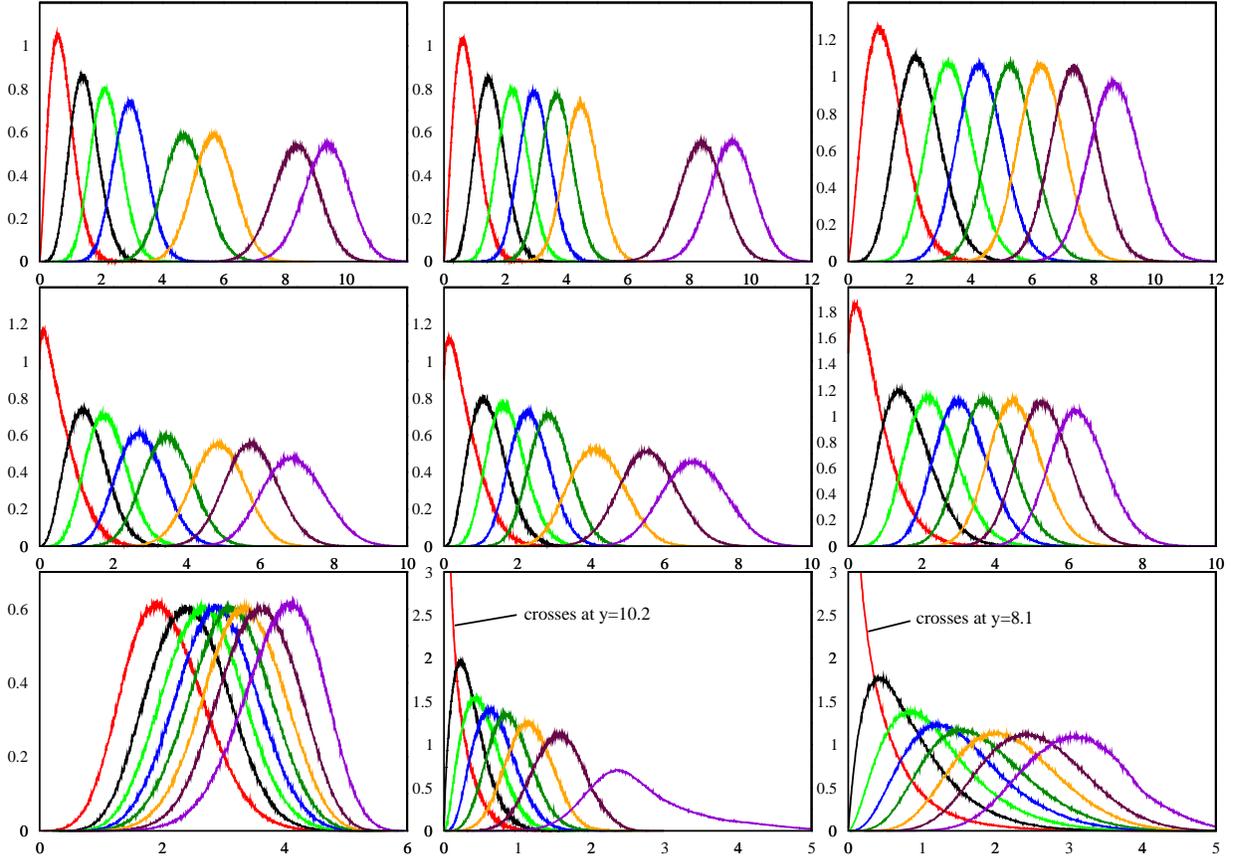}
 \caption{Normalized PGB mass-squared distributions for the SU(3) models discussed in the text. The models types and flavors are the same as in subsection~\ref{subsec1}. The columns and rows represent model type and flavor respectively. For example the rightmost graph in the middle row represent results model MODEL IIC.}
 \label{fig:su3}
\end{figure}

The PGB mass distributions are presented in Fig~\ref{fig:su3}. We also present the percentage occurrences of various phases in Table~\ref{tablesu3}.

The first thing to notice is that models IA and IB (and IIA and IIB) do not have the same PGB distributions, unlike the SU(2) case. This is hint about an asymmetry in the minimum and maximum structures of a given model with respect to $\Lambda\to-\Lambda$. As it was the case in SU(2),  we see that the probability of occurrence of a light PGB mass is considerable in models of type II and III compared to type I where T-invariance is {\em not} imposed a priori. These probabilities are larger than the corresponding ones in SU(2). This is understandable because in SU(3) imposing T-invariance constrains more elements out of the total possibilities compared to SU(2). In type III models this effect is much more emphasized, except for IIIA, where by construction the minimization is always achieved by $S=I$. 
As a minor note we would like to point out that the PGB mass distributions for models IA show an almost-good SU(2) symmetry since the PGB masses seem to form clusters with cardinality $1+3+2+2$ coinciding with the SU(2) decomposition of the adjoint of SU(3), that is $\rm{8=1+3+2+2^{*}}$. For the other cases such an observation is not possible and SU(2) is badly broken.
Thus the behavior of PGB mass distributions is similar to what we have obtained in SU(2) except that in SU(3) an almost vanishing PGB mass is more probable.

\begin{table}[h]
\begin{tabular}{|c||c|c|c|}
\hline
MODEL & CPC & PCP & CPV\\
\hline
IA & 0.0 & 0.0 & 100.0 \\
\hline
IB & 0.0 & 0.0 & 100.0 \\
\hline
IC & 0.0 & 0.0 & 100.0\\
\hline
IIA & 86.3 & 0.0 & 13.7 \\
\hline
IIB & 84.0 & 0.0 & 16.0\\
\hline
IIC & 77.4 & 0.0  & 22.6 \\
\hline
IIIA & 100.0 & 0.0 & 0.0\\
\hline
IIIB & 41.3 & 44.7 & 14.0 \\
\hline
IIIC & 62.0 & 31.4 & 6.6 \\
\hline
\end{tabular}
\caption{The occurrence percentages of different phases for the SU(3) models studied. Model types and flavors are the same
as in subsection~\ref{subsec1}. All numbers have a statistical error of 0.1 percent.}
\label{tablesu3}
\end{table}
We see from Table~\ref{tablesu3} that for all models of type I the solutions are in the CPV phase. This makes sense because in these models T-invariance is not originally imposed. What is illuminating is that for models of type II there are no PCP phases. In reality, we have observed a tiny fraction of models resulting in the PCP phase. But this was less than the statistical error rendering the result compatible with 0. This observation implies that the PCP phases are not living in the bulk of the space of most general models with original T-invariance. This is expected since PCP phases are represented by O(N) type orbits that is not connected to the identity element and thus occur with more contrivedness than the CPC cases which are proper O(N) subgroups. For type IIIA since the minimization is always achieved by $S=I$ all models yield a CPC solution. The PCP phases show themselves in models IIIB and IIIC. We therefor conclude that in SU(3) CPC phases are more common and PCP phases occur with respectable frequency only in models where there are more constraints than just T-invariance. The benchmark model studied in \cite{kenestiame, kenadam2} is such an example.

Therefor as a general conclusion we simply state that within the context of vacuum alignment and the possible subsequent spontaneous
breaking of the CP symmetry there is considerable possibility for a light PGB mass if in the original hamiltonian T-invariance is imposed. As a speculation we would like to argue that it seems as if a similar mechanism to that of the Peccei-Quinn model operates, of course only in a statistical sense \footnote{To arrive at this statistical conclusion we made use of random models sampled from regions  $\left[0,1\right]$,  $\left[-1,0\right]$ and  $\left[-1,1\right]$. One can object to our treatment in view of the fact that since the origin is part of the sampling space some elements are probably much smaller than the others. This is a legitimate criticism even though the number of hierarchically separated elements is low with high probability. Nevertheless
we have also studied the mentioned models with sampling regions of the form $\left[1,2\right]$,  $\left[-2,-1\right]$ and  $\left[-2,-1\right] \bigcup \left[1,2\right]$ where no hierarchy whatsoever is present. The structure of the PGB mass distributions are the same albeit the distributions are lower at zero mass, meaning that the effect is present but less emphasized. So hierarchy within the elements
of $\Lambda_{ab}$ is important. But we have already seen this concept in our discussion; there models of type III which were models of type II with only diagonal entries yielded larger probabilities for small PGB masses.}.

\section{Conclusions and Future Directions}

We studied in detail the vacuum alignment problem in models where a global chiral symmetry first spontaneously breaks to its vectorial subgroup and later is explicitly broken by a perturbation of four-Fermi type. We exposed different CP behavior classes and transitions
between those classes, presented an explicit solution for two flavor case, SU(2). We have also shown with statistical evidence that there is an increased chance for a light pseudo-Goldstone boson if the perturbation has an ab initio CP invariance.
 
The most obvious thing to pursue next is to be more energetic for finding an analytical treatment for the three flavor case SU(3), since this is the case with the lowest number of flavors where all of the possible CP violation classes are present. A complete list of extrema like that of SU(2) is what one desires, however there is no clear path to follow for this. A recipe to discriminate the CP structure of
the minimum is actually enough to understand the behavior of the model. As we have mentioned there are simplified models where an
exact treatment is possible and a diagonal $\Lambda_{ab}$ seems to be the best starting point.

Another avenue which is of interest is to allow mass terms. In a model generalized in this manner the energy function to minimize becomes

\be
E(W)=-\alpha{\rm Tr}\left[MW+W^{\dagger}M^{\dagger}\right]-\beta\Lambda_{ab}{\rm Tr}\left[W^{\dagger}\lambda^{a}W\lambda^{b}\right]\;.
\ee

\noindent In this work we have studied $\alpha=0$ case. The other extreme, $\beta=0$, is a well known system with a research literature too wide to enumerate here. See however recent interesting results in \cite{creutz1,creutz2}. It will be interesting how an interplay between these two terms can change things. Works along these lines are in progress.

\acknowledgments{}
I thank Kenneth Lane and Adam Martin for numerous useful correspondence and to Teoman Turgut and Burak Kaynak for many stimulating discussions on the topic. I have also benefited from previous collaboration with Kenneth Lane and Estia Eichten. The Monte-Carlo analysis mentioned in the text is performed on Kassandra, the HPC cluster of Department of Physics at Bo\~{g}azi\c{c}i University.

\end{document}